\documentclass[aps, prb, twocolumn, showpacs, floatfix]{revtex4-1}
\usepackage{amsmath, amsfonts, amssymb, mathbbol}

\usepackage{graphicx,color}
\usepackage{hyperref}

\usepackage[normalem]{ulem}

\newcommand{\titlestr}{Magneto-optical response of graphene: probing substrate interactions}

\begin{document}

\title{\titlestr}

\newcommand{\itptuw}{}

\author{L.A.~Chizhova}
\email{larisa.chizhova@tuwien.ac.at}
\affiliation{Institute for Theoretical Physics, Vienna
  University of Technology, Wiedner Hauptstra{\ss}e 8-10, 1040 Vienna,
  Austria, EU}

\author{J.~Burgd\"orfer}
\affiliation{Institute for Theoretical Physics, Vienna
  University of Technology, Wiedner Hauptstra{\ss}e 8-10, 1040 Vienna,
  Austria, EU}

\author{F.~Libisch}
\affiliation{Institute for Theoretical Physics, Vienna
  University of Technology, Wiedner Hauptstra{\ss}e 8-10, 1040 Vienna,
  Austria, EU}

\date{\today}

\begin{abstract}
Magneto-optical transitions between Landau levels can provide precise
spectroscopic information on the electronic structure and excitation spectra of graphene,
enabling probes of substrate and many-body effects.
We calculate the magneto-optical conductivity of large-size graphene
flakes using a tight-binding approach. Our method allows us to
directly compare the magneto-optical response of an isolated graphene
flake with one aligned on hexagonal boron nitride giving rise
to a periodic superlattice potential. The substrate interaction induces
band gaps away from the Dirac point. In the presence of a perpendicular magnetic field
Landau-level like structures emerge from these zero-field band gaps.
The energy dependence of these satellite structures is, however,
not easily accessible by conventional probes of the density of states
by varying the back-gate voltage. Here we propose the magneto-optical
probing of the superlattice perturbed spectrum.
Our simulation includes magneto-excitonic effects in first-order
perturbation theory. Our approach yields a quantitative explanation of
recently observed Landau-level dependent renormalizations of the Fermi
velocity.
\end{abstract}

\pacs{73.22.Pr,  71.70.Di, 81.05.ue, 71.70.-d}

\maketitle

\section{Introduction}
\label{section1}

The progress in fabrication of well-characterized graphene structures
led to the availability of graphene devices with mobilities as high as
$6 \cdot 10^5$ cm$^2$/Vs \cite{Tombros11}.
As transport properties are strongly influenced by disorder,
the use of suitable substrates such as hexagonal boron nitride (hBN)
are key to substantially
reduce the amount of substrate-induced bulk disorder compared with
conventional SiO$_2$, dramatically improving the electronic and transport
properties of graphene \cite{Wang13}. These substrates, however,
can substantially modify the physical properties of graphene \cite{Dean10, Yankowitz12}.
For example, perfectly aligned graphene on
hBN features a moir\'e pattern of 14 nm periodicity due to a small lattice
mismatch between the two materials \cite{Yankowitz12}.
The superlattice potential leads to the
opening of a gap at the Dirac cone and to the folding of the Brillouin zone.
Intersections between the original cone and its back-folded replica
give rise to additional band gaps energetically above and below the
main Dirac cone \cite{Yankowitz12, Ponomarenko13}.
In the presence of a magnetic field perpendicular to the graphene,
secondary satellite Landau levels emanate near these band gaps.
Moreover, the large real-space superlattice unit cell allows
observing the Hofstadter butterfly \cite{Hofstadter76} at laboratory
accessible field strength \cite{Gorbachev14, Hunt13}.

Recent experiments \cite{Gorbachev14, Hunt13, Dean10} probing the
satellite structures and the Hofstadter butterfly of graphene on hBN in a
magnetic field have employed a varying back gate voltage ($V_{\mathrm{bg}}$)
to control the effective Fermi level $E_F$ in the graphene sheet.
Such a measurement does not, however, provide unambiguous information
on the density of states:  through quantum capacitance effects and deviations
from the idealized linear density of states of graphene (caused,
e.g., by the satellite structures), the relation between energy and
applied back gate voltage may deviate substantially from
expectations \cite{Chizhova14}. An attractive experimental alternative
is magneto-spectroscopy \cite{Chen14, Jiang07, Sadowski06}.
This technique provides information on the energy difference between
the ground state and the particle-hole excitation and, thus,
between the different Landau levels. Effective Fermi velocities can be
extracted from a fit of Dirac Landau levels to the
appropriate transitions. In the presence of
a substrate induced moir\'e potential, the magneto-optical signal can
also provide information on the satellites and the evolution of
associated Landau levels.
The observed shift of the inter-Landau level transitions of graphene
on different substrates can be described by a
renormalized Fermi velocity and attributed to many-body effects  \cite{Chen14, Jiang07}.
Velocity renormalization was also measured in magneto-Raman scattering
experiments \cite{Faugeras14, Neumann15}.

In this work, we determine the magneto-optical conductivity of large ($140
\times 120 \mbox{ nm}^2$) graphene flakes where the influence of residual edge
effects is suppressed by imposing a Berry-Mondragon potential
\cite{Berry87} at the boundary (see also Refs.~\cite{Libisch10,
  Chizhova14}).  We evaluate the optical conductivity by calculating
the dipole transition between the different eigenstates of the flake,
which we obtain within the third-nearest neighbor tight-binding
approximation.  We compare the behavior of a pristine graphene flake
with a graphene flake aligned with hBN represented by a superlattice
potential \cite{Chizhova14}.  Absorption lines associated with the
optical transitions between Landau levels of the Dirac fermions show
the expected square root dependence on the magnetic field.  For
graphene on hBN, we observe a Hofstadter butterfly on top of each
transition line. The satellite structures above and below the Dirac
cone, which are the signatures of the Brillouin zone folding in
graphene aligned with hBN, are shown to contribute to the optical conductivity: our
tight-binding approach predicts additional structures in the optical
conductivity that evolve linearly with magnetic field, opening a pathway
for optical characterization of graphene-substrate interactions.  Furthermore,
we calculate the shift of the inter-Landau levels transition lines of
pristine graphene in the magnetic field due to an attraction of
excited electron and hole, i.e.~magneto-excitons, on a tight-binding
level.  We solve a two-body problem and evaluate the direct Coulomb
electron-hole interaction within first-order perturbation theory. The
resulting velocity renormalization agrees well with the experimental
data \cite{Chen14, Jiang07} and with complementary theoretical approaches for bulk
graphene \cite{Iyengar07, Shizuya10}. One advantage of the present tight-binding
based approach is the inclusion of spatially varying
long-range potentials due to a substrate.

\section{Magneto-optical response of pristine graphene}
\label{section2}
\begin{figure*}[tb]
 \centering
    \includegraphics[width=1.0\textwidth]{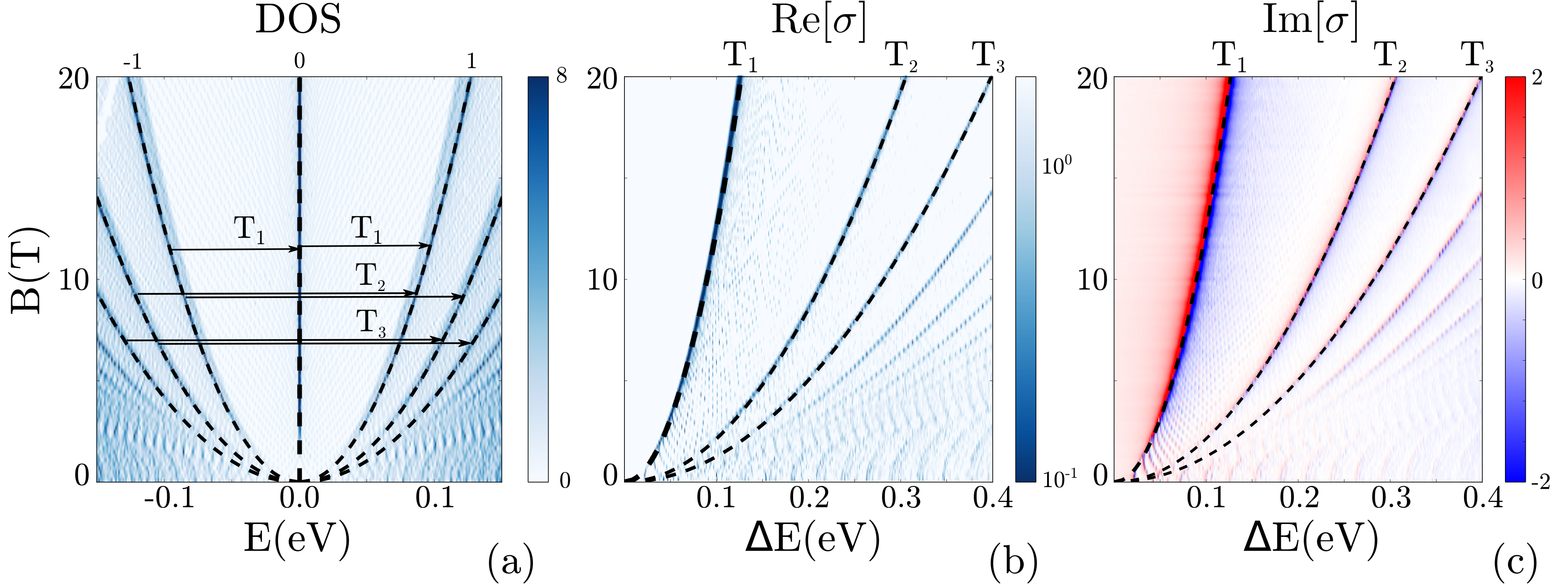}
    \caption{(a) The density of states of a 140x120 nm$^2$ pristine graphene flake as a function of
    energy and magnetic field. Black dashed curves denote the Landau levels of massless
    Dirac fermions (Eq.(\ref{eq:lanlev})).
    (b) Real (absorptive) and (c) imaginary (dispersive) part of the optical conductivity $\sigma$ (color scale) of the flake
    (evaluated using Eq.(\ref{eq:opt_cond2})) as a function of the excitation energy $\Delta E =\hbar \omega$ and magnetic field.
    The black dashed curves are the analytical prediction for
    the optical inter-Landau levels transitions [see Eq.(\ref{eq:trans})].}
\label{fig:opt_cond_ideal}
\end{figure*}

The tight-binding (TB) method is very useful for simulating large-scale structures when more sophisticated techniques
relying on periodic boundary conditions and small supercell sizes
[e.g., density functional theory (DFT)] prove computationally challenging due to
the presence of a magnetic field or additional large-scale potential
variations.
While for small magnetic fields (i.e.~$B\approx0$)
the density of states (DOS) [see Fig.\ref{fig:opt_cond_ideal}(a)]
is dominated by size quantization and properties of the flake boundaries,
for increasing field, Landau levels begin to
emerge resembling the behavior of bulk graphene.
The transition from the linear DOS of bulk graphene at $B=0$ to the Landau
level regime is governed by the magnetic length $l_B \simeq 25.5/\sqrt{B[\mathrm{T}]}[\mathrm{nm}]$.
For flakes \cite{Libisch10} Landau levels appear when $l_B$ becomes
smaller than the flake diameter $D$.
The Landau level $N$ for massless Dirac fermions of graphene satisfies \cite{Rabi28}
\begin{equation}
E^D_N = \mathrm{sgn}(N) v_F \sqrt{2 e\hbar N B},
\label{eq:lanlev}
\end{equation}
which coincides with the high density regions in the calculated DOS
[see black dashed curves in Fig.\ref{fig:opt_cond_ideal}(a)].  Here
$v_F$ is the Fermi velocity of the Dirac dispersion $E_D = \pm v_F
|\vec{p}|$.  For the present set of tight-binding parameters
\cite{Libisch12} the Fermi velocity of pristine graphene is $v^0_{F} =
0.78 \cdot 10^6$ m/s.  This value of Fermi velocity determined from
DFT \cite{Reich02}, however, does not include the many-body effects
seen in experiments \cite{Gillen10, Deacon07, Sprinkle09} and
corrections discussed below.

Landau levels in graphene were observed in transport
experiments and were measured by changing the back gate voltage
$V_{\mathrm{bg}}$.
The back gate voltage changes the number of charge carries $n=\alpha V_{\mathrm{bg}}$
in graphene and, therefore, the Fermi energy
\begin{equation}
E_F = \hbar v_F \sqrt{\pi n}
\label{eq:rescale}
\end{equation}
through capacitive coupling between the back gate and the graphene sheet,
where $\alpha$, the so-called lever arm, is a measure for the coupling
strength.
Combining Eqs.~(\ref{eq:lanlev}) and (\ref{eq:rescale}) yields straight
lines for Landau levels in the V$_{\mathrm{bg}}$-B plane in
contrast to the square-root behavior in the E-B plane for Landau levels except of $N=0$.
By contrast,
magneto-spectroscopy allows for direct probing of Eq.(\ref{eq:lanlev}) by optical transitions between
Landau levels. Distortions of the linear DOS by traps or localized charge states especially at the device edges,
which influence the $V_{\mathrm{bg}}$ coupling, are largely eliminated.

The selection rule for optically allowed inter-Landau levels transitions
is $\Delta N = |N_f| - |N_i| = \pm 1$, where $N_{f}$($N_{i}$) is the final (initial)
Landau level quantum number \cite{Moon13}.
Deviations of the realistic graphene bandstructure from the ideal Dirac
cone such as trigonal warping or the perturbation by the interaction with the
substrate will break this selection rule, and transitions with $|\Delta N| > 1$
will become allowed \cite{Moon13}.
However, the probability of these transitions remains small. The main transition
lines visible in the measured infrared absorption spectra \cite{Chen14, Jiang07}
are $T_1 = E^D_1 - E^D_{0}$, $T_2 = E^D_2 - E^D_{-1}$
and $T_3 = E^D_3 - E^D_{-2}$. The analytical forms of the transition energies,
which correspond to the peak positions in the associated absorption
spectrum, follow from Eq.(\ref{eq:lanlev}) as
\begin{equation}
\begin{aligned}
    T_1 &= \sqrt{2|e|\hbar v_F^2B}, \\
    T_2 &= \sqrt{2|e|\hbar v_F^2B} (\sqrt{2} + 1), \\
    T_3 &= \sqrt{2|e|\hbar v_F^2B} (\sqrt{3} + \sqrt{2}).
\end{aligned}
\label{eq:trans}
\end{equation}
Employing the eigenstates and eigenenergies of the tight-binding
calculation for the graphene flake, it is straightforward to obtain the
optical dipole transitions between the different eigenstates due to the
coupling to the electromagnetic field of the infrared laser. The associated
optical conductivity is given in the length gauge by \cite{Moon13}
\begin{equation}
\sigma(\omega) = \frac{i e^2}{\hbar S} \sum \frac{f(\epsilon_a)-f(\epsilon_b)}{\epsilon_a -
\epsilon_b - \hbar \omega + i \eta} (\epsilon_a - \epsilon_b) |\langle a | \hat{x} | b \rangle|^2,
\label{eq:opt_cond2}
\end{equation}
where $\omega$ is a photon frequency of the infrared laser, $S$ is the area of the flake,
and $\epsilon_a$ ($\epsilon_b$) are the eigenenergies
of the flake corresponding to the eigenstate $|a\rangle$ ($|b\rangle$). The summation
indices $a$ and $b$ extend over all the eigenstates of the flake in the selected
energy window. The matrix element $\langle a | \hat{x} | b \rangle$
gives the transition dipole moment between the two eigenstates.
We use the Fermi distribution $f(\epsilon_{a,b})$ at zero temperature
with the chemical potential located at the Dirac point.
The difference of Fermi distributions, $f(\epsilon_{a})-f(\epsilon_{b})$, is non-zero for
transitions between occupied and unoccupied states corresponding to
particle-hole excitation. Zeros in the denominator signify resonant absorption
of photons with $\hbar \omega = \epsilon_a - \epsilon_b$. In order to avoid
numerical instabilities we have introduced a line broadening of $\eta = 0.1$ meV
of the discrete eigenstates of the flake, which is, however,
small compared to the physical width of the coarse-grained DOS and, thus,
of no consequence for the numerical results.

Real and imaginary parts of the magneto-optical conductivity calculated using Eq.(\ref{eq:opt_cond2})
for the pristine graphene flake [Fig.\ref{fig:opt_cond_ideal}(b, c)] confirm
the dominance of the optically allowed transitions of the Dirac cone
also for the finite-size graphene flake.
The maxima of the (absorptive) real part of the optical conductivity as well as the nodal lines of the (dispersive)
imaginary part agree well with the analytical prediction (Eq.(\ref{eq:trans})) using the Fermi velocity $v^0_F$
[see black dashed curves in Fig.\ref{fig:opt_cond_ideal}(b, c)].
The observed magneto-optical transition lines $T_1$, $T_2$, $T_3$ etc.~are narrow
with widths of the Landau levels  in the DOS, of about $\approx 2$meV [Fig.\ref{fig:opt_cond_ideal}(a)].
Dipole-forbidden transitions with $\Delta N > 1$ are not visible in Fig.\ref{fig:opt_cond_ideal}(b, c)
because of their much weaker oscillator strength.

\section{Magneto-optical response of graphene on \lowercase{h}BN}
\label{section3}

The tight-binding approximation allows us to study the magneto-optical
conductivity of more complex structures with broken periodicity
or large periodic supercells. Specifically, perfectly
aligned graphene on hexagonal boron nitride (hBN)
features a 14 nm periodic moir\'e pattern due to the small
lattice mismatch \cite{Hunt13, Gorbachev14, Yankowitz12}.
An effective potential reproducing the main features of the measured DOS
is of the form \cite{Chizhova14}
\begin{equation}
V_{\mathrm{BN}}(\vec{r}) = W(\vec{r}) \cdot \sigma_z + V(\vec{r}) \cdot  \mathbb{I} + W_{\mathrm{MB}}(B, \vec{r}) \cdot \sigma_z,
\label{eq:pot}
\end{equation}
where $\sigma_z$ is the Pauli matrix. This potential has the
periodicity of the moir\'e pattern.
The first term $W(\vec{r}) \cdot \sigma_z$
is responsible for the breaking of the sublattice symmetry and for
opening of a band gap of $\Delta=14$ meV
near the Dirac point. The presence of a gap modifies the Dirac Landau levels to
\begin{equation}
E^D_N(B) = \left\{
  \begin{array}{lr}
    \text{sgn}(N)\sqrt{2|e|\hbar v_F^2|N|B+(\Delta/2)^2} : N \neq 0\\
    \pm \Delta/2 : N = 0^{\pm},
  \end{array}
\right.
\label{eq:lanlev_mod}
\end{equation}
The two zeroth Landau levels to which we assign the quantum numbers $0^+$ and $0^-$
corresponding to energies $\pm \Delta/2$  emerge as a result of the
valley splitting. The valley degeneracy of the other Landau levels is not lifted.
The second term in Eq.(\ref{eq:pot}) is a smoothly varying potential accounting for
stronger binding in regions where
one sublattice of graphene is on top of boron and another sublattice
is in the middle of the BN hexagon. This term accounts
for the increased electron-hole asymmetry beyond the third nearest-neighbor
tight-binding approximation. In addition, $V(\vec{r})$ may modify the Fermi velocity.
For the parameters of $V(\vec{r})$ considered, the Fermi velocity of graphene
subject to superlattice potential of hBN decreases to $v^{\mathrm{BN}}_F = 0.76 \cdot 10^6$ m/s, which is
slightly smaller than that of pristine graphene $v^0_F$. Finally, the last
term $W_{\mathrm{MB}}(B, \vec{r}) \cdot \sigma_z$
in Eq.(\ref{eq:pot}) describes, on a phenomenological level, magnetic-field dependent
many-body effects and accounts for the magnetic-field induced valley splitting.
The lifting of the four-fold degenerate (two-fold for valley and two-fold for spin) zeroth
Landau level was observed in graphene on
SiO$_2$ \cite{Jiang07b, Giesbers09}, hBN \cite{Young12, Hunt13}
and in suspended graphene \cite{Andrei09}.
In graphene on hBN \cite{Young12}, the valley splitting was experimentally found to be
linearly proportional to the magnetic field, a result which is still unexplained
by theory. Two alternative theoretical methods have been previously proposed
to explain this observation. (i) One model, based on the continuous Dirac model
in the Hartree-Fock approximation,
predicts a $\sqrt{B}$ scaling of the valley splitting \cite{Abanin13, Nomura06}.
While this scaling differs from the linear scaling, the resulting numerical values
for the valley splitting resembles those of the experiment within the range of
investigated magnetic field strength.
(ii) The second model, based on the effect
of lattice distortion and the interaction with the substrate,
predicts linear scaling of valley splitting with $B$
but underestimates the strength of the splitting by an order of magnitude  \cite{Fuchs07}.
In the present work, we account for the valley splitting within the
single-electron model by applying the phenomenological
potential $W_{\mathrm{MB}}(B, \vec{r})$,
which depends linearly on the magnetic field with the spatially averaged value
$\bar{W}_{\mathrm{MB}}(B) = 8 B[\mathrm{T}]$ meV/T
consistent with experiment \cite{Young12}.
The spatial variation of the potential $W_{\mathrm{MB}}$
follows the Gaussian shape of  $V(\vec{r})$, i.e.~${W}_{\mathrm{MB}}(B, \vec{r})=\bar{W}_{\mathrm{MB}}(B) \cdot V(\vec{r})/ \left<V(\vec{r})\right>$.
(The Zeeman spin splitting, also proportional to $B$, is neglected in the
following).

\begin{figure*}[tb]
 \centering
    \includegraphics[width=1.0\textwidth]{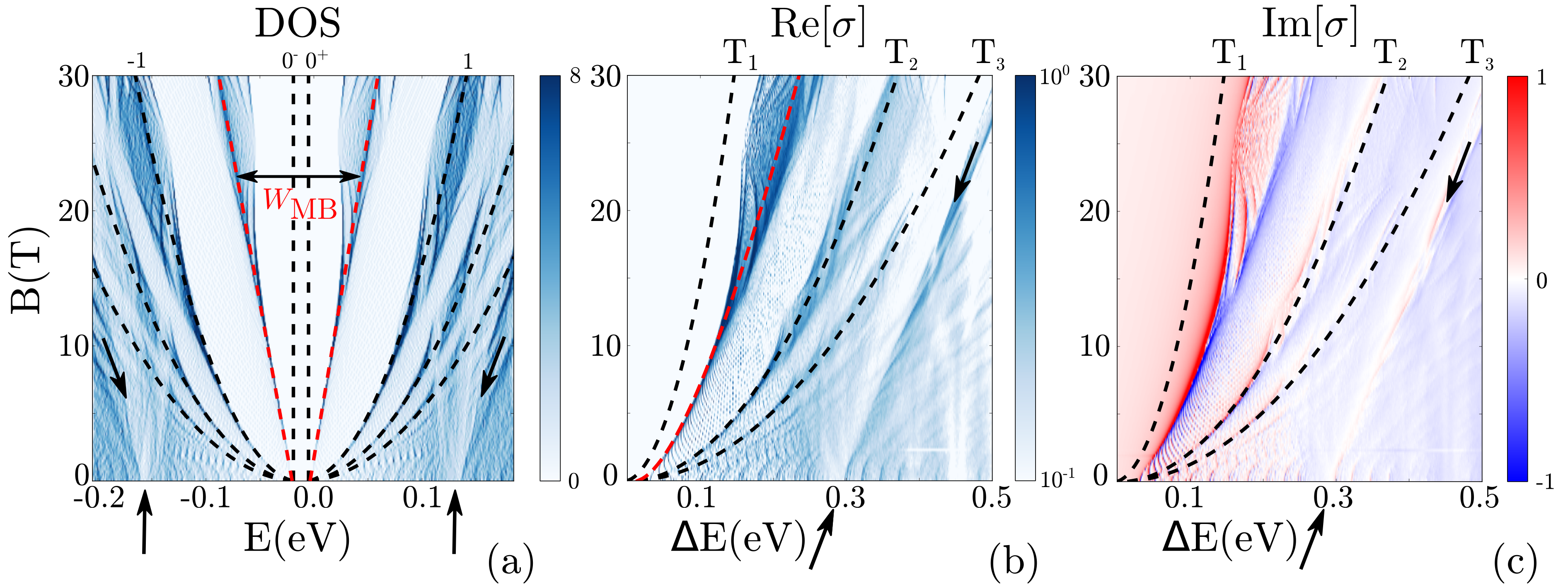}
    \caption{Same as Fig.\ref{fig:opt_cond_ideal} but for a 140x120 nm$^2$ graphene flake
    deposited on hexagonal boron nitride (hBN) giving rise to a periodic superlattice potential with period
    14 nm period.
    The black dashed curves in (a) correspond to the Landau levels of Dirac fermions with a
    finite mass [Eq.(\ref{eq:lanlev_mod})].
    The red dashed lines indicate the magnetic field dependent valley splitting in the zeroth Landau level
    due to the additional phenomenological potential $W_{\mathrm{MB}}(B, \vec{r})$.
    The satellites above and below the Dirac point at $B=0$ and their magnetic field evolution
    are marked by arrows. In (b) and (c)
    black dashed curves denote the optical transitions $T_1$, $T_2$ and $T_3$,
     predicted by the difference between the corresponding Landau levels for
     massive Dirac fermions (Eq.(\ref{eq:lanlev_mod})).
    The red dashed trace in (b) corresponds to the $T_1$ transition when the magnetic field
    dependent valley splitting of the zeroth Landau level is included.
    The optical transitions between Landau levels at $B=0$ of
    satellites and their B-field evolution are marked by arrows (compare to Fig.\ref{fig:opt_cond_ideal}(b,c)).}
\label{fig:opt_cond}
\end{figure*}

The DOS of the graphene flake on hBN [Fig.\ref{fig:opt_cond}(a)]
as well as the optical conductivity [Fig.\ref{fig:opt_cond}(b, c)]
feature the square-root scaling for the $N \neq 0$ Landau levels $E^D_N \sim \sqrt{B}$
of massive Dirac fermions given by Eq.(\ref{eq:lanlev_mod})
[see black dashed curves in Fig.\ref{fig:opt_cond}(a)]
for magnetic fields up to $B \approx 10$ T. By contrast,
the linear magnetic field dependence of the zeroth Landau level
is determined by the effective many-body potential
$W_{\mathrm{MB}}(\vec{r})$. [If $W_{\mathrm{MB}}=0$,
the zeroth Landau levels, i.e.~$0^{+}$ and $0^{-}$, coincide with the magnetic
field independent vertical black dashed curves in Fig.\ref{fig:opt_cond}(a)].
The position of the absorption lines in the calculated optical conductivity agrees
with the energy difference between Landau levels (Eq.(\ref{eq:lanlev_mod}))
for $N \neq 0$ using Fermi velocity $v^{\mathrm{BN}}_F$. For transitions involving the zeroth ($N=0$)
level, e.g.~the $T_1$ transition, we have to include in Eq.(\ref{eq:lanlev_mod}) the linear magnetic field dependence of the
$0^-$ Landau level [see red dashed curve in Fig.\ref{fig:opt_cond}(b)].
The sensitivity of the $T_1$ transition to the many-body valley (and spin)
splitting in the zeroth Landau level offers an alternative
to activation gaps measurements.\cite{Young12}

At higher magnetic fields, the Hofstadter butterfly \cite{Hofstadter76}
becomes visible as "diamond"-like structures on top of each Landau level
in the DOS [see Fig.\ref{fig:opt_cond}(a) at $B \gtrsim 10$ T].
This phenomenon arises due to the competition
between two length scales: the magnetic length $l_B= 25.5/\sqrt{B[\mathrm{T}]}[\mathrm{nm}]$ and the scale of the hBN
superlattice with a period of 14 nm. At $B=25$ T, i.e.~when
the magnetic flux through the superlattice unit cell is equal
to the magnetic flux quantum, the DOS features the diamonds with
the largest extent.
The optical conductivity, likewise, features a Hofstadter butterfly
modifying the absorption lines [see Fig.\ref{fig:opt_cond}(b, c) at $B \gtrsim 10$ T].
The energetic width of the diamond (FWHM) of the $T_1$ transition at 25 T is 20 meV
corresponding to the size of the first
Landau level diamond obtained from the density of states
[Fig.\ref{fig:opt_cond}(a)].

Another superlattice induced effect is the presence of satellites above and below the Dirac point
near $E=-0.15$ eV and $E=0.14$ eV at $B=0$ [see arrows in Fig.\ref{fig:opt_cond}(a)],
which arise due to the avoided crossings between the Dirac cone and its replica.
The corresponding Landau levels of the satellites evolve linearly
with magnetic field in the DOS.
These satellite structures appear also in the optical conductivity as an absorption line
starting at $\Delta E = 0.3$ eV at $B=0$ and linearly increasing with $B$
[marked by arrows in Fig.\ref{fig:opt_cond}(b,c)].
Its slope corresponds to the slope of the satellite Landau levels.
The pronounced difference in the magnetic field evolution of Landau
levels of the satellites ($\sim B$) and the
main Dirac cone ($\sim \sqrt{B}$) is not obvious when measured as a function of
$V_{\mathrm{bg}}$.
Optical transition spectroscopy appears as an attractive alternative
for disentangling
Dirac-like and Schr\"odinger-like dispersions in the bandstructure.
Magneto-optical experiments for graphene on hBN \cite{Chen14}
have, up to now, not focused on the satellite structures induced by
the moir\'e pattern. While challenging, the experimental observations
of the satellites and their magnetic field evolution would yield
valuable insight into the influence of the moir\'e potential on the bandstructure.

\section{Magneto-excitonic shift of the optical inter-Landau levels absorption lines}
\label{section4}

In the single-particle picture we used so far, the calculated optical conductivity
yields the position of absorption lines $T_i$ in agreement
with analytical predictions [Eq.(\ref{eq:trans})] based on the difference between
the corresponding Landau levels of Dirac fermions [Eq.(\ref{eq:lanlev}) and Eq.(\ref{eq:lanlev_mod})]
In the experiment \cite{Chen14, Jiang07}, however,
one observes deviations which can be conveniently parameterized in terms of
transition-line dependent Fermi velocities $v^{T_i}_F$ in the $\Delta E \sim \sqrt{B}$
relation (Eq.(\ref{eq:lanlev_mod})).
Their origin are obviously excitonic interactions correcting for particle-hole excitations
absent in addition spectroscopy when varying the back-gate potential $V_{\mathrm{bg}}$.

The formation of magneto-excitons \cite{Iyengar07, Bychkov08} by optical
excitations between the Landau levels is a many-body effect.
To lowest order the excitonic wave function is a product of an electron and a hole wave function:
\begin{equation}
\Psi^{\mathrm{exc}}_{NM}(\vec{r}_{\mathrm{el}}, \vec{r}_{\mathrm{h}}) =
\psi_N (\vec{r}_{\mathrm{el}}) \psi_M (\vec{r}_{\mathrm{h}}),
\end{equation}
where $\vec{r}_{\mathrm{el}}$ ($\vec{r}_{\mathrm{h}}$) is an electron
(hole) coordinate, $ \psi_N $ and $ \psi_M $ correspond to the wave functions
of the $N^{th}$ and $M^{th}$ Landau levels undergoing optical transitions.
The corresponding Dirac equation reads
\begin{equation}
\hat{H} \Psi^{\mathrm{exc}}_{NM}(\vec{r}_{\mathrm{el}}, \vec{r}_{\mathrm{h}}) =
E \Psi^{\mathrm{exc}}_{NM}(\vec{r}_{\mathrm{el}}, \vec{r}_{\mathrm{h}}), \\
\end{equation}
where
\begin{equation}
\hat{H} = \sum_{i=el,h} v_F \vec{\sigma} \cdot (\vec{p}_i - e \vec{A}) -
\frac{e^2}{4\pi \varepsilon_0 \varepsilon \left| \vec{r}_{\mathrm{el}} - \vec{r}_{\mathrm{h}} \right| },
\label{eq:Exc}
\end{equation}
$\varepsilon$ is a dielectric constant and $\varepsilon_0$ is the
permittivity of the vacuum.  We use the value of $\varepsilon = 5$
found for graphene on SiO$_2$ within the random phase approximation
(RPA) \cite{Alicea06} and also for graphene on hBN
\cite{Faugeras14}. In the absence of the Coulomb interaction [second
  term in $\hat{H}$, Eq.(\ref{eq:Exc})] the excitonic energy coincides
with Eq.(\ref{eq:trans}), i.e.~the single-particle energy
difference. The Coulomb interaction changes the transition energies
and leads to the observed energy shift due to particle-hole
attraction. Magneto-excitonic effects were studied in detail for the
two-dimensional electron gas \cite{Haug:PS, Kallin84} (2DEG) and for
graphene \cite{Iyengar07, Bychkov08, Shizuya10}.  In particular, it
was shown that the excitation energy consists of several
contributions: (i) the (non-interacting) single-particle exciton
energy $\Delta E = E^D_{N} - E^D_{N'}$ (Eq.(\ref{eq:lanlev_mod}));
(ii) the direct Coulomb two-particle interaction between the particle
and the hole; (iii) the annihilation and creation of electron-hole
pairs at different points of the Brillouin zone; and (iv) the exchange
interaction. The direct Coulomb term (ii) is negative and gives rise
to the excitonic binding. Therefore, the optical transition energy is
reduced relative to the estimate in the single-particle picture.
Contributions (iii) and (iv) provide positive higher-order
corrections, slightly reducing the Coulomb attraction. Since this
effect is already, to a certain extent, empirically accounted for by the dielectric response
of the medium $\varepsilon$, we neglect an explicit treatment of these
terms.

\begin{figure*}[tb]
 \centering
    \includegraphics[width=1.0\textwidth]{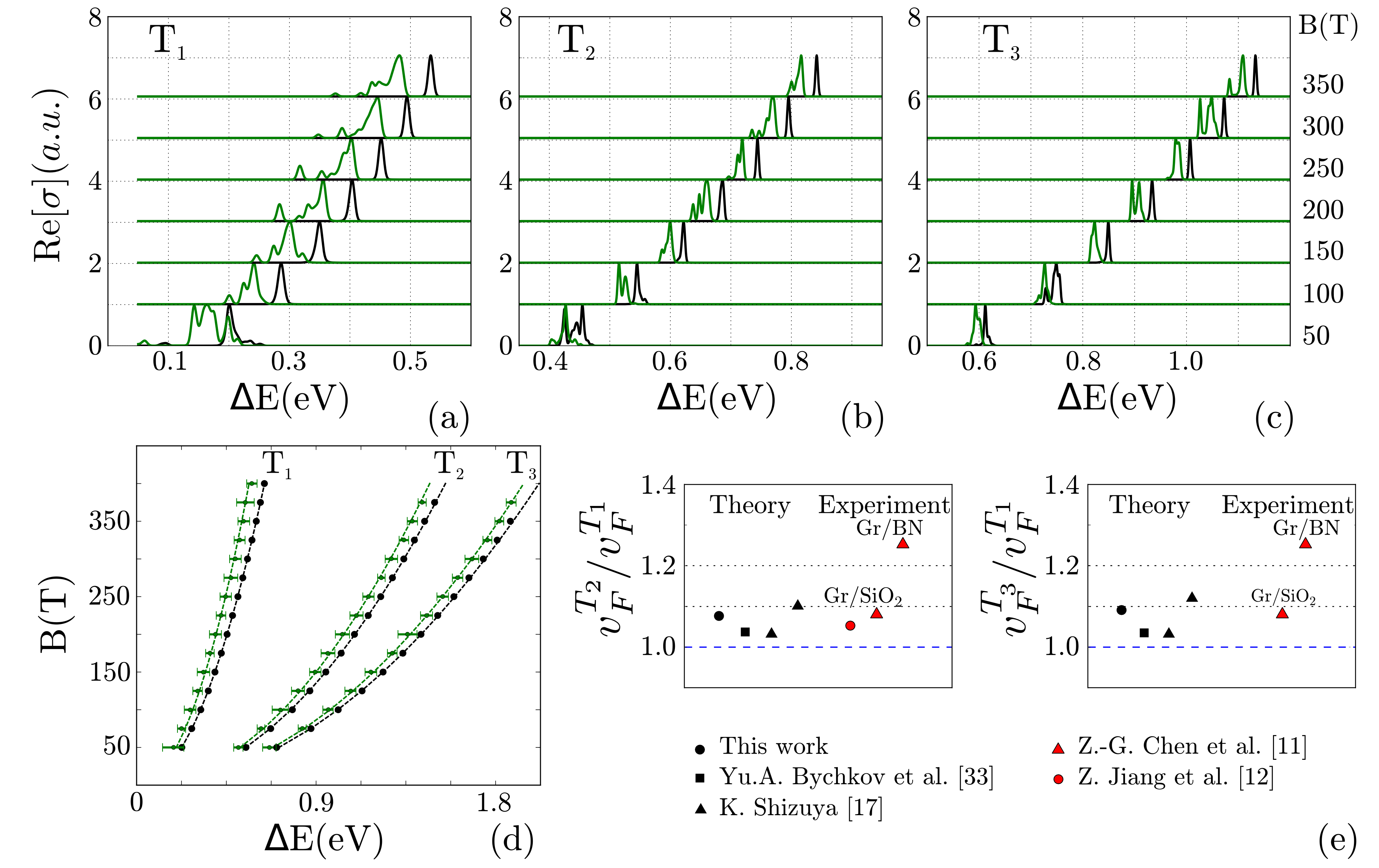}
    \caption{(a, b, c) Real part of the optical conductivity [renormalized to its maximum] of the
    transition lines T$_1$, T$_2$ and T$_3$ as a function of photon energy.
    Conductance traces at different magnetic fields are shifted vertically.
    Black traces correspond to single-particle
    transitions, green traces correspond to optical transitions taking into
    account the direct Coulomb electron-hole interactions from
    first-order perturbation theory.
    (d) The position (dots) and width (bars) of spectral lines extracted from
    the first and second moment of the
    single-particle (black dots) and excitonic excitation spectra (green dots).
    The black dashed curves correspond to the analytically predicted
    Landau level transitions given by Eq.(\ref{eq:trans}) with a Fermi velocity
    $v^0_{F} = 0.78 \cdot 10^6$ m/s.
    The green dashed curves correspond to the fit of the lines with Eq.(\ref{eq:trans}), however,
    with renormalized Fermi velocities: $v^{T_1}_{F} = 0.69 \cdot 10^6$ m/s,
    $v^{T_2}_{F} = 0.745 \cdot 10^6$ m/s and $v^{T_3}_{F} = 0.755 \cdot 10^6$ m/s.
    (e) The ratio between Fermi velocities of T$_{2}$ (right) and T$_3$ (left) transitions to
    the Fermi velocity of T$_1$: theory versus experiment.}
\label{fig:excitons}
\end{figure*}

Within the tight-binding approximation applied to graphene dots,
it is possible to treat magneto-excitonic effects.  For comparison with the experiment
we include the dielectric environment for graphene on a SiO$_2$ substrate by taking into
account the dielectric constant $\varepsilon=5$ when evaluating the Coulomb interaction term.
We treat the direct Coulomb electron-hole
interaction in first-order perturbation theory. Consider an optical
transition between the two Landau levels $N$ and $M$.
Each of the Landau levels comprises a large number of eigenstates of the flake
[see Fig.(\ref{fig:opt_cond_ideal}(a))].
A photoexcited state can be written as a superposition of flake eigenstates
with energies in a vicinity of the optically excited
Landau levels:
\begin{equation}
\Psi^{exc}_{NM}(\vec{r}_{\mathrm{el}}, \vec{r}_{\mathrm{h}})=
\sum_{a, b} C_{ab} \phi^N_a(\vec{r}_{\mathrm{el}}) \phi^M_b(\vec{r}_{\mathrm{h}}).
\label{eq:wf}
\end{equation}
The expansion coefficients $C_{ab}$ can be obtained by inserting Eq.(\ref{eq:wf}) into the Dirac equation
Eq.(\ref{eq:Exc}):
\begin{equation}
(\epsilon^N_a - \epsilon^M_b) C_{ab} - \sum_{a',b'} V^{ab}_{a'b'} C_{a'b'} = \epsilon C_{ab},
\label{eq:BSE}
\end{equation}
with $\epsilon^N_a$ ($\epsilon^M_b$) being eigenenergies of the flake in the vicinity of
$N^{th}$ ($M^{th}$) Landau level.
The Coulomb matrix element is
\begin{equation}
\begin{aligned}
V^{ab}_{a'b'} &=\frac{e^2}{4\pi \epsilon_0 \epsilon}\\
                          &\iint d\vec{r}_{\mathrm{el}} d\vec{r}_{\mathrm{h}}
\frac{\phi^{*N}_{a'}(\vec{r}_{\mathrm{el}})\phi^{*M}_{b'}(\vec{r}_{\mathrm{h}})
\phi^{N}_{a}(\vec{r}_{\mathrm{el}})\phi^{M}_{b}(\vec{r}_{\mathrm{h}})}
{\left| \vec{r}_{\mathrm{el}} - \vec{r}_{\mathrm{h}}\right|}
\end{aligned}
\label{eq:matrix_el}
\end{equation}
Eq.(\ref{eq:BSE}) can be viewed as the Bethe-–Salpeter equation (BSE)
\cite{Louie08} adapted for a finite-size graphene flake.  The Coulomb
integral Eq.(\ref{eq:matrix_el}) gives rise to the exciton binding
energy.  The full solution of Eq.(\ref{eq:BSE}) is numerically
challenging as the size of the subspace coupled by the Coulomb matrix
is given by the product of the number of particle and hole states for
each Landau level. 

For each Landau level involved in the transition, we consider the
energetically closest $N_S$ eigenstates of the flake. Given the size
of our model system, we find that using $N_S\approx$ 30$-$40 (depending on
the magnetic field and the Landau level index) is a good compromise between
accuracy and numerical effort: the matrix dimensions of $V^{ab}_{a'b'}$ are
given by $N_S^2 \times N_S^2 \approx 10^6$, were the evaluation of each matrix element
requires a double spatial integral with a Coulomb kernel. To keep the
problem numerically manageable, we (i) restrict ourselves to
first-order perturbation theory \cite{Landau:QM}, i.e.~we include only
the diagonal Coulomb matrix elements $V^{ab}_{ab}$ in
Eq.(\ref{eq:BSE}) and (ii) reduce the size of the graphene flake to
$24 \times 24 \mbox{ nm}^2$, to speed up the calculation of each
matrix element. As we verified numerically, the diagonal elements
$V^{ab}_{ab}$, indeed, dominate. For each electron-hole pair $| a, b
\rangle$ the corresponding first-order energy correction to the
noninteracting electron-hole transition energies
$\epsilon_{a,b}^{(0)}=(\epsilon_a - \epsilon_b)$ is given by the
diagonal matrix element
\begin{equation}
\epsilon_{a,b}^{(1)}=V^{ab}_{ab} =\frac{e^2}{4\pi \epsilon_0 \epsilon}
                          \iint d\vec{r}_{\mathrm{el}} d\vec{r}_{\mathrm{h}}
\frac{|\phi^{N}_{a}(\vec{r}_{\mathrm{el}})|^2 |\phi^{M}_{b}(\vec{r}_{\mathrm{h}})|^2}
{\left| \vec{r}_{\mathrm{el}} - \vec{r}_{\mathrm{h}}\right|}.
\label{eq:matrix_el_diag}
\end{equation}
The binding energy of the $| a, b \rangle$ exciton is in
the diagonal approximation given by $\epsilon_{a,b} = \epsilon_{a,b}^{(0)} -
\epsilon_{a,b}^{(1)}$.

The optical conductivity of each electron-hole pair, which is related
to the dipole matrix element between the electron $|a\rangle$ and hole
$|b\rangle$ eigenstates, is calculated using Eq.(\ref{eq:opt_cond2}),
however, now with the excitonic energy contribution
included. First-order perturbation theory results in a pronounced
shift of the T$_1$, T$_2$ and T$_3$ transition lines in the direction
of smaller photon energies, when comparing the noninteracting single
particle picture [black traces in Fig.\ref{fig:excitons}(a,b,c)] with
the two-particle correction [green traces in
  Fig.\ref{fig:excitons}(a,b,c)].  The position and the width of the
spectral lines is determined by their first and second
moments. Neglecting the excitonic effect the transition energies
[black dots in Fig.\ref{fig:excitons}(d)] agree well with the
analytical prediction for inter-Landau levels optical transitions
given by Eq.(\ref{eq:trans}) [black curves in
  Fig.\ref{fig:excitons}(d)] with a Fermi velocity $v^0_{F} = 0.78
\cdot 10^6$ m/s determined by our choice of tight-binding parameters
for pristine graphene. In contrast, the two-particle excitonic
corrections shift the lines to lower transition energies $\Delta E$
[green data points in Fig.\ref{fig:excitons}(d)].  A fit of these
transition lines to the analytic prediction $\Delta E \sim \sqrt{B}$
yields now line-specific Fermi velocities.  In particular, for T$_1$
the Fermi velocity is $v^{T_1}_{F} = 0.69 \cdot 10^6$ m/s; for T$_2$,
$v^{T_2}_{F} = 0.745 \cdot 10^6$ m/s and for T$_3$, $v^{T_3}_{F} =
0.755 \cdot 10^6$ m/s. Such a velocity renormalization, also seen in
the experiment \cite{Chen14, Jiang07}, clearly reflects excitonic
effects. To distinguish effects of the Landau-level specific velocity
renormalization from other bandstructure effects that uniformly affect
all transitions, we consider the ratio of the renormalized Fermi
velocities of T$_{2,3}$ transitions to that of T$_1$. We find
reasonable agreement with the experiments \cite{Chen14, Jiang07}
measuring optical inter-Landau levels transitions for graphene on
SiO$_2$ and also with alternative theoretical approaches for bulk graphene
\cite{Bychkov08, Shizuya10} [Fig.\ref{fig:excitons}(e)]. The present
tight binding ansatz allows for including long-range disorder,
such as puddles observed in graphene on SiO$_2$. We account for the
effects of charge puddles by applying a smooth disorder potential with
an amplitude of 25 meV (50 meV) and a correlation length of 10 nm
(5nm) \cite{Martin07}. Disorder only slightly modifies the
theoretically calculated line shifts and velocity renormalization.  The
resulting optical inter-Landau levels transition lines
remain well-defined in the presence of disorder in agreement with magneto-optical
experiments for graphene on SiO$_2$.

  However, the
magneto-optical measurements of graphene on hBN \cite{Chen14} notably
differ from other measurements and theoretical predictions [see right
  most point in Fig.\ref{fig:excitons}(e)]. This mismatch arises from
a substantially ($\approx 30\%$) lower experimental value of
$v_F^{T_1}$ than theoretical predictions and was attributed to
many-body effects influencing the zeroth Landau level
\cite{Chen14}. Corrections to the dielectric response of the material
(i.e., the effective $\varepsilon$) should affect all transitions
similarly, and thus cannot explain the observed large ratio
$v_F^{T_2}/v_F^{T_1}$. One effect large enough to explain these
findings is the substantial splitting of the four-fold degenerate
Landau levels of pristine graphene, as found for graphene on hBN in,
e.g., quantum capacitance measurements \cite{Gorbachev14}. However,
any such splitting should shift the $T_1$ line to higher photon
energies [as seen in Fig.\ref{fig:opt_cond}(b,c)], i.e.~in the
opposite direction to that observed in the experiment. Clearly, future
magneto-optical experiments for graphene on hBN are called for to shed
more light on this puzzle.

One remark should be added regarding the large values of the magnetic
fields used in the simulations to map out Landau levels.  Since we
consider a flake of a smaller size than in the previous section, the
Landau levels emerge at higher magnetic fields. In particular, the
Landau level $N$ is formed as soon as $2\sqrt{2N}l_B$ is smaller than
the smallest dimension of the flake.  For example, the $N=3$ Landau
level participating in the $T_3$ optical transition becomes
distinguishable only at $B \gtrsim 30$ T for the small flake used to
calculate magneto-excitonic corrections. Clearly, the optical response
of the larger flakes used in experiment can still be extracted from
the present calculation by extrapolating the square root behavior of
the optical transitions with transition-dependent Fermi velocity to
lower fields.

\section{Conclusion}
\label{sec:conclusion}
We have simulated the optical properties of graphene flakes with and
without moir\'e potential for aligned graphene on hBN using the
tight-binding approximation.  Our simulations show that the
magneto-optical response allows probing the satellites due to
Brillouin zone folding and, unlike probing by back gate voltage, to
clearly distinguish energy levels that scale linearly with $B$ from
those that feature a square-root scaling, $\sim \sqrt{B}$.  We have
also shown that excitonic effects can be included in a
tight-binding description. We validate our predictions for a
Landau-level specific renormalization of the Fermi velocity by
comparing to experimental data for graphene on SiO$_2$, opening a pathway
towards the description of excitonic effects in larger structures.
For graphene on hexagonal boron nitride, current experimental data
does not fully agree with theoretical predictions, calling for further
experimental and theoretical studies.

\section*{Acknowledgments}
\label{sec:acknowledgments}
We gratefully acknowledge support from the doctoral college Solids4Fun (FWF), as well as by ViCom (SFB 041-ViCom).
Calculations were performed on the Vienna Scientific Cluster.


\begin{thebibliography}{10}

\bibitem{Tombros11}
N.~Tombros, A.~Veligura, J.~Junesch, J.J.~van den Berg, P.J.~Zomer, M.~Wojtaszek, I.J.~Vera Marun, H.T.~Jonkman, and B.J.~van Wees
\newblock {\em J.~Appl.~Phys.}, {\bf 109}, 093702 (2011).

\bibitem{Wang13}
L.~Wang, I.~Meric, P.Y.~Huang, Q.~Gao, Y.~Gao, H.~Tran, T.~Taniguchi, K.~Watanabe, L.M.~Campos, D.A.~Muller, J.~Guo, P.~Kim, J.~Hone, K.L.~Shepard, and C.R.~Dean
\newblock {\em Science}, {\bf 342}, 614 (2013).

\bibitem{Dean10}
C.~R. Dean, A.~F. Young, I.~Meric, C.~Lee, L.~Wang, S.~Sorgenfrei, K.~Watanabe,
  T.~Taniguchi, P.~Kim, K.~L. Shepard, and J.~Hone.
\newblock {\em Nature Nano.} {\bf 5}, 722 (2010).

\bibitem{Yankowitz12}
Matthew Yankowitz, Jiamin Xue, Daniel Cormode, Javier~D. Sanchez-Yamagishi,
  K.~Watanabe, T.~Taniguchi, Pablo Jarillo-Herrero, Philippe Jacquod, and
  Brian~J. LeRoy.
\newblock {\em Nature Phys.} {\bf 8}, 382 (2012).

\bibitem{Ponomarenko13}
L.A.~Ponomarenko, R.V.~Corbachev, G.L.~Yu et al.
\newblock {\em Nature} {\bf 497}, 594 (2014).

\bibitem{Hofstadter76}
Douglas~R. Hofstadter.
\newblock {\em Phys.~Rev.~B}, {\bf 14}, 2239 (1976).

\bibitem{Gorbachev14}
G.L.~Yu, et al.
\newblock {\em Nature Phys.} {\bf 10}, 525 (2014).

\bibitem{Hunt13}
B.~Hunt, J.~D. Sanchez-Yamagishi, A.~F. Young, M.~Yankowitz, B.~J. LeRoy,
  K.~Watanabe, T.~Taniguchi, P.~Moon, M.~Koshino, P.~Jarillo-Herrero, and R.~C.
  Ashoori.
\newblock {\em Science} {\bf 340}, 1427 (2013).

\bibitem{Chizhova14}
L.A.~Chizhova, F.~Libisch, and J.~Burgd\"orfer.
\newblock {\rm Phys.~Rev.~B}, {\bf 90}, 165404 (2014).

\bibitem{Sadowski06}
M.L.~Sadowski, G.~Martinez, M.~Potemski, C.~Berger and W.A.~de Heer
\newblock {\rm Phys.~Rev.~Lett.}, {\bf 97}, 266405 (2006).

\bibitem{Chen14}
Z.-G. Chen, Z. Shi, W. Yang, X. Lu, Y. Lai, H. Yan, F. Wang, G. Zhang, and Z. Li,
\newblock {\em Nature Comm.} {\bf 5}, 4416 (2014).

\bibitem{Jiang07}
Z.~Jiang et.al
\newblock {\em Phys.~Rev.~Lett.~} {\bf 98}, 197403 (2007).

\bibitem{Faugeras14}
C.~Faugeras, S.~Berciaud, P.~Leszczynski, Y.~Henni, K.~Nogajewski, M.~Orlita, T.~Taniguchi, K.~Watanabe, C.~Forsythe, P.~Kim, R.~Jalil, A.K.~Geim, D.M.~Basko, and M.~Potemski
\newblock {\em Phys.~Rev.~Lett.~} {\bf 114}, 126804 (2015).

\bibitem{Neumann15}
C.~Neumann, S.~Reichardt, M.~Dr\"{o}geler, B.~Terr\'{e}s, K.~Watanabe, T.Taniguchi, B.~Beschoten, S.V.~Rotkin, and C.~Stampfer.
\newblock {\em Nano Lett.~} {\bf 15}, 1547 (2015)

\bibitem{Berry87}
M.V.~Berry and R.J.~Mondragon.
\newblock {\em Proc.~R.~Soc.~Lond., Ser.~A} {\bf 412}, 53 (1987).

\bibitem{Libisch10}
F.~Libisch, S.~Rotter, J.~G\"{u}ttinger, C.~Stampfer, and J.~Burgd\"{o}rfer.
\newblock {\em Phys.~Rev.~B} {\bf 81}, 245411 (2010).

\bibitem{Shizuya10}
K.~Shizuya
\newblock {\em Phys.~Rev.~B} {\bf 81}, 075407 (2010)

\bibitem{Iyengar07}
A.~Iyengar, J.~Wang, H.A.~Fertig and L.Brey.
\newblock {\em Phys.~Rev.~B} {\bf 75}, 125430 (2007).

\bibitem{Rabi28}
I.~I. Rabi.
\newblock {\em Z.~Phys.~A} {\bf 49}, 507 (1928).

\bibitem{Libisch12}
F.~Libisch, S.~Rotter, and J.~Burgd\"orfer.
\newblock {\rm New~Jour.~of~Phys.~}, {\bf 14}, 123006 (2012).

\bibitem{Reich02}
S.~Reich, J.~Maultzsch, C.~Thomsen, and P.~Ordej\'on
\newblock {\em Phys.~Rev.~B} {\bf 66}, 035412 (2002).

\bibitem{Sprinkle09}
M.~Sprinkle et.al
\newblock {\em Phys.~Rev.~Lett.} {\bf  103}, 226803 (2009)

\bibitem{Gillen10}
R.~Gillen and J.~Robertson
\newblock {\em Phys.~Rev.~B} {\bf 82}, 125406 (2010)

\bibitem{Deacon07}
R.~S.~Deacon, K.C.~Chuang, R.~J.~Nicholas, K.~S.~Novoselov and  A.~K.~Geim
\newblock {\em Phys.~Rev.~B} {\bf 76}, 081406 (2007)

\bibitem{Moon13}
P.~Moon and M.~Koshino
\newblock {\em Phys.~Rev.~B} {\bf 88}, 241412(R) (2013)

\bibitem{Giesbers09}
A.~J.~M.~Giesbers , L.A.~Ponomarenko, K.S.~Novoselov, A.K.~Geim, M.I.~Katsnelson, J.C.~Maan, and U.~Zeitler.
\newblock {\em Phys.~Rev.~B} {\bf 80}, 201403(R) (2009).

\bibitem{Jiang07b}
Z.~Jiang, Y.~Zhang, H.L.~Stormer, and P.~Kim
\newblock {\em Phys.~Rev.~Lett.~} {\bf 99}, 106802 (2007).

\bibitem{Young12}
A.~F. Young, C.~R. Dean, L.~Wang, H.~Ren, P.~Cadden-Zimansky, K.~Watanabe,
  T.~Taniguchi, J.~Hone, K.~L. Shepard, and P.~Kim.
\newblock {\em Nature Phys.} {\bf 8}, 550 (2012).

\bibitem{Andrei09}
X.~Du, I.~Skachko, F.~Duerr, A.~Luican, and E.Y.~Andrei
\newblock {\em Nature} {\bf 462}, 192 (2009).

\bibitem{Abanin13}
D.A.~Abanin, B.E.~Feldman, A.~Yacoby and B.I.~Halperin
\newblock {\em Phys.~Rev.~B} {\bf 88}, 115407 (2013).

\bibitem{Nomura06}
K.~Nomura, and A.H.~MacDonald
\newblock {\em Phys.~Rev.~Lett.~} {\bf 96}, 256602 (2006).

\bibitem{Fuchs07}
J.-N.~Fuchs and P.~Lederer
\newblock {\em Phys.~Rev.~Lett.~} {\bf 98}, 016803 (2007).

\bibitem{Bychkov08}
Yu.A.~Bychkov and G.~Martinez.
\newblock {\em Phys.~Rev.~B} {\bf 77}, 125417 (2008).

\bibitem{Alicea06}
J.~Alicea, and M.P.A.~Fisher
\newblock {\em Phys.~Rev.~B} {\bf 74}, 075422 (2006).

\bibitem{Kallin84}
C.~Kallin and B.I.~Halperin
\newblock {\em Phys.~Rev.~B} {\bf 30}, 5655 (1984)

\bibitem{Haug:PS}
H.~Haug
\newblock {\em {Quantum theory of the optical and electronic properties of semiconductors}},
\newblock World Scientific Publishing Co.~Pte.~Ltd., 3nd edition (1994).

\bibitem{Louie08}
L.~Yang, M.L.~Cohen and S.G.~Louie
\newblock {\em Phys.~Rev.~Lett.} {\bf  101}, 186401 (2008)

\bibitem{Landau:QM}
L.D.~Landau and E.M.~Lifshitz.
\newblock {\em {Quantum Mechanics. Non-relativistic theory}}, volume~3.
\newblock Pergamon Press, London, 2nd edition (1965).

\bibitem{Martin07}
J.~Martin et.al
\newblock {\em Nature Phys.} {\bf 4}, 144 (2007).

\end{thebibliography}
\end{document}